\documentstyle[twocolumn,aps,pre,graphics]{revtex}
\begin{document}
\bibliographystyle{prsty} 

\title{Pattern Dynamics of Rayleigh--B{\'e}nard 
convective rolls and weakly
  segregated diblock copolymers}

\author{Jacob J. Christensen$^1$ and Alan J. Bray$^2$}

\address{ $^1$Institute of Physics and Astronomy, 
University of Aarhus,
  DK-8000 Aarhus C, Denmark \\ $^2$Department of Physics 
and Astronomy, The
  University of Manchester, \\ Manchester, M13 9PL, UK}

%\date{\today}
\address{\em (\today)}
%\maketitle
\address{
\centering{
\medskip\em
\begin{minipage}{14cm}
{}~~~We consider the pattern dynamics of the lamellar phases observed in
  Rayleigh--B{\'e}nard convection, as described by the Swift--Hohenberg
  equation, and in the weak segregation regime of diblock copolymers. Both
  numerical and analytical investigations show that the dynamical growth of
  the characteristic length scale in both systems is described by the same
  growth exponents, thus suggesting that both systems are members of the
  same universality class. 
\end{minipage}
}}

%47.54.+r  Pattern selection; pattern formation
%47.27.Te  Convection and heat transfer
%64.75.+g  Solubility, segregation, and mixing; phase separation
%83.10.Nn  Polymer dynamics
\maketitle

\narrowtext

\section{Introduction}
The study of the dynamics of pattern formation in systems far from
equilibrium encompasses examples from both physics, chemistry and biology
\cite{CrossHohenberg94}. Despite completely different physical origins some
systems exhibit identical morphologies and pattern dynamics, and may be
perceived as members of the same universality class.

\begin{figure}[hb]
  \begin{picture}(1.0,0.75)
    \put(-0.25,-0.95){\resizebox{1.3\columnwidth}{!}{\includegraphics{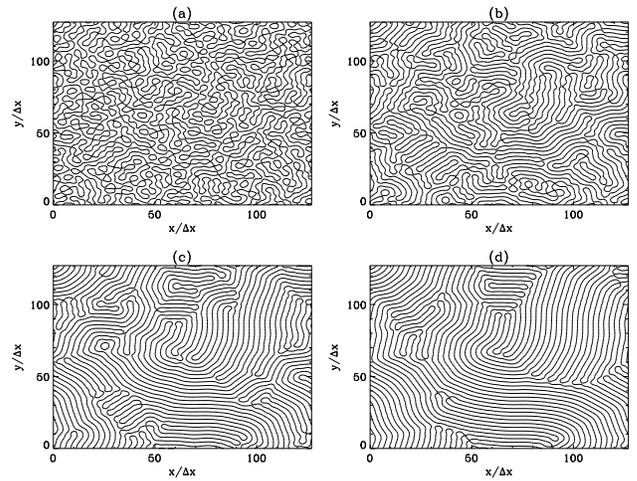}}}
  \end{picture}
  \caption{The coarsening process. The figure shows 
    snapshots of the domain configurations in the diblock polymer system
    shortly after the quench from the disordered to the bistable phase (a)
    and at increasingly later times (b)-(d). The pictures are contour plots
    of $128 \times 128$ systems where the contours are defined by
    $\phi(x,y,t)=0$. The order parameter field, $\phi(x,y,t)$, was obtained
    by numerical integration of Eq.(\ref{DC}) at zero thermal noise.
    Simulations of the Swift--Hohenberg system [Eq.(\ref{SH})] produces
    domain configurations which morphologically are indistinguishable from
    those here presented for the diblock copolymer system.}
  \label{fig:coarse}
\end{figure}

In this paper we consider the pattern dynamics of two morphologically
identical systems, namely Rayleigh--B{\'e}nard convective rolls and weakly
segregated diblock copolymers. At short times after a quench from the
uniform stable phase to the unstable phase both systems develop a
labyrinthine domain morphology consisting of rolls (or lamellae) of a
well--defined width $w$. Initially the rolls are randomly oriented, but as
time increases they locally align up in parallel thereby creating an
increasingly ordered pattern [Fig.\ref{fig:coarse}].  We have investigated
the dynamics of this coarsening process by numerical integration of the
appropriate Langevin equations and by analytical considerations. Both
approaches agree that the characteristic length scale of the systems scales
dynamically with growth exponents which are common to both systems, thereby
suggesting that the pattern dynamics of Rayleigh--B{\'e}nard convection and
diblock copolymers belong to the same universality class.

The observed ordering phenomenon is driven by two mechanisms, namely
interface relaxation and defect annihilation. The effect of the former
mechanism can in a defect--free system be calculated by considering the
speed at which a modulated interface relaxes to its (straight) ground
state. Specifically we apply the projection operator method \cite{Elder92}
developed for interface relaxation in the Rayleigh--B{\'e}nard system to
the same problem in diblock copolymers, thus providing a systematic
treatment of both systems.  Furthermore we show how the application of a
general approach to interface relaxation recently developed by one of the
authors \cite{Bray98} leads to the same result for the Rayleigh--B{\'e}nard
system.

This paper is organized as follows: In Sec.\ref{sec:Models} we introduce
the two models we study.  Our numerical work is presented and discussed in
Sec.\ref{sec:Simulations}. Sec.\ref{sec:Theory} contains the theoretical
considerations including a brief review of the projection operator method,
and Sec.\ref{sec:Discussion} concludes with a summary and discussion.

%%%%%%%%%%%%%%%%%%%% FIGURE %%%%%%%%%%%%%%%%%%%%%%%%
%%%%%%%%%%%%%%%%%%%%%%%%%%%%%%%%%%%%%%%%%%%%%%%%%%%%

%
%--------------- M O D E L S ---------------------------
%

\section{Models} \label{sec:Models}
In the Rayleigh--B{\'e}nard system a simple fluid is confined between two
horizontal plates which are heated from below, and for values of the
Rayleigh number, $R$, larger than a critical value, $R_c$, an instability
occurs which transform the uniform state to a state consisting of spatially
periodic convective rolls.  Near the onset of the convective instability
the free energy functional, $F$, of the Rayleigh--B{\'e}nard system is, in
dimensionless variables, well approximated by the form
\begin{equation}
  \label{FSH}
  F[\phi] = \int 
d^2r\,(-\phi[\epsilon-(k_0^2+\nabla^2)^2]\phi/2+\phi^4/4
),
\end{equation}
deduced by Swift and Hohenberg (SH) \cite{SH77}. Here the scalar
order--parameter field $\phi=\phi(x,y,t)$ is related to the local vertical
fluid velocity, $\epsilon=(R-R_c)/R_c$ is the reduced Rayleigh number which
acts as the control parameter of the system, and $k_0$ is the wavenumber
corresponding to the period, $\lambda=2w$, of the modulated structure,
i.e., $k_0=\pi/w$.

For small $\epsilon$ the order parameter field is locally well described by
a single mode approximation, $\phi({\bf r}) \sim \cos({\bf k}\cdot{\bf r})$, where $\bf k$ is perpendicular to the
orientation of the rolls, and as $\epsilon \rightarrow 0$ this
approximation is exact \cite{PomeauMan79}.  Minimizing the free energy
Eq.(\ref{FSH}) in the single mode approximation yields $k=k_0$ as the
selected wavenumber of the steady state. As customary we use $k_0=1$.  The
equation of motion for $\phi$ is given by the Langevin equation
$\partial_t\phi=-\delta F[\phi]/\delta\phi+\zeta$, where $F$ is the above
free energy and $\zeta=\zeta({\bf r},t)$ is thermal noise correlated as
$\langle \zeta({\bf r},t) \zeta({\bf r}',t')\rangle =2A\delta({\bf r}-{\bf
  r}')\delta(t-t')$, where $A$ parameterizes the strength of the thermal
fluctuations. Thus the Swift--Hohenberg equation reads
\begin{equation}
  \label{SH}
  \partial_t\phi = 
\epsilon\phi-(k_0^2+\nabla^2)^2\phi-\phi^3+\zeta. 
\end{equation}

A diblock copolymer (DC) is a linear chain molecule joined together by two
strings of equal length of e.g. $A$ and $B$ monomers.  The polymerization
index, $N$, is thus $N=N_A+N_B$ where $N_A=N_B$ are the numbers of $A$ and
$B$ monomers, respectively. Above the critical temperature $T_c$, $A$ and
$B$ mix, whereas below $T_c$ the two sequences are incompatible and the
copolymer melt undergoes phase separation.  However, spinodal decomposition
\cite{Bray94} cannot continue indefinitely because of the chemical bond
between the sequences. As a result the phase separation occurs on a length
scale bounded above by the length of a stretched polymer chain (typically
less than 1 micrometer) where banded domains of $A$--rich and $B$-rich
regions alternate in the final equilibrium state. The free energy of a
diblock copolymer melt below $T_c$ is given (also in dimensionless
variables) by a modified Cahn--Hilliard free energy functional
\cite{LiuGolden89}
\begin{eqnarray}
  \label{FDC}
  F[\phi] &=& \int d^dr\, [f(\phi)+(1/2)(\nabla\phi)^2] 
\nonumber \\
&+&(\Gamma/2) \int
  d^dr\,d^dr'\,\phi({\bf r})G({\bf r},{\bf r}')\phi({\bf 
r}'),
\end{eqnarray}
where $\phi({\bf r},t)=\phi_A({\bf r},t)-\phi_B({\bf r},t)$ is the local
concentration difference between the $A$ and $B$ monomers,
$f(\phi)=-\phi^2/2+\phi^4/4$, is the bulk free energy density, and $\Gamma$
is a control parameter inversely proportional to the square of the
polymerization index, $\Gamma \sim 1/N^2$. Finally, the Green's function,
$G({\bf r},{\bf r}')$, in the second integral is defined by the Poisson
equation, $\nabla^2G({\bf r},{\bf r}')=-\delta({\bf r}-{\bf r}')$.  The
order parameter for this system is a conserved quantity, thus the
appropriate Langevin equation for the time evolution of $\phi$ subsequent
to a quench from the disordered to the bistable phase, is
$\partial_t\phi=\nabla^2\delta F[\phi]/\delta\phi+\zeta$, or inserting
Eq.(\ref{FDC}),
\begin{equation}
  \label{DC1}
  \partial_t\phi=\nabla^2(-\phi+\phi^3-\nabla^2\phi)-\Gamma\phi+\zeta,
\end{equation}
where the noise $\zeta$, representing the effect of thermal fluctuations,
has the correlations $\langle \zeta({\bf r},t) \zeta({\bf r}',t')\rangle
=-2A\nabla^2\delta({\bf r}-{\bf r}')\delta(t-t')$. For $\Gamma$ just below
the critical value, $\Gamma_c=1/4$, Eq.(\ref{DC1}) describes the dynamics
of weakly segregated lamellar domains with a well--defined width
$w=\pi/k_0$, where $k_0=\Gamma^{1/4}$ is the wavenumber which minimizes the
free energy [Eq.(\ref{FDC})] in a single mode approximation
\cite{Christensen}.

The diblock copolymer equation [Eq.(\ref{DC1})] can conveniently be
rewritten in a form resembling the Swift--Hohenberg equation
[Eq.(\ref{SH})],
\begin{equation}
   \label{DC}
    \partial_t\phi = 
\epsilon\phi-(1/2+\nabla^2)^2\phi+\nabla^2\phi^3+\zeta,
\end{equation}
where $\epsilon=\Gamma_c-\Gamma$.  Linearizing in Fourier space about
$\phi=0$ we find, in both Eq.(\ref{SH}) and Eq.(\ref{DC}), that
fluctuations, $\delta\phi_k$, in the order parameter decay exponentially,
$\delta\phi_k(t) =\delta\phi_k(0)\exp[-\omega_kt]$ with rate
$\omega_k=(\alpha-k^2)^2 -\epsilon$, where $\alpha=1,1/2$ for the SH and DC
system, respectively. Thus both systems have a band of wavevectors, $k_- <
k < k_+$, $k_\pm=\sqrt{\alpha\pm\sqrt{\epsilon}}$, for which the uniform
state is unstable. In the nomenclature of Cross and Hohenberg
\cite{CrossHohenberg94} this means that both systems are stationary
periodic, or Type $I_s$.

%
%------------- S I M U L A T I O N S ------------------------
%

\section{Simulations} \label{sec:Simulations}
We have solved the SH and DC equations numerically using a finite
difference scheme on two dimensional lattices of size $512 \times 512$,
with periodic boundary conditions.  Numerical algorithms for the
spatio--temporal evolution of both systems were obtained by replacing, in
Eq.(\ref{SH}) and Eq.(\ref{DC}), $\partial_t\phi({\bf r},t)$ by
$(\phi_{ij}^{n+1}-\phi_{ij}^n)/\Delta t$, and $\nabla^2\phi({\bf r},t)$ by
the following discretized Laplacian
\begin{equation}
  \label{La}
  \nabla^2\phi_{ij} =\frac{1}{(\Delta
    x)^2}\left[\frac{2}{3}\sum_{(NN)}+\frac{1}{6}\sum_{(NNN)}
  -\frac{10}{3} \right]\phi_{ij},
\end{equation}
which includes contributions from both nearest neighbours (NN) and
next--nearest neighbours (NNN).  Here the indices $(i,j)$ represent the
coordinates $(x,y)$ and the index $n$ represents time.  Connection to
absolute time and spatial coordinates is established through the
relationships $t=n\Delta t$ and ${\bf r}=(i\hat{x}+j\hat{y})\Delta x$.  The
specific choice of coefficients in Eq.(\ref{La}) ensures that the
Laplacian, in Fourier space, is isotropic to second order in $k^2$, i.e.,
the form of the Fourier transform, $\Gamma_{\bf k}$, of Eq.(\ref{La}), is
$\Gamma_{\bf k}=-k^2+const.\times(\Delta x)^2k^4 + O((\Delta x \,k^2)^3)$.
For the diblock copolymer system the fluctuation--dissipation relation for
the discrete equation can be maintained by generating two independent
Gaussian variables $\nu^{(1)}_{ij}(n)$, $\nu^{(2)}_{ij}(n)$ with zero mean
and correlations $ \langle \nu^{(a)}_{ij}(m)\nu^{(b)}_{kl}(n) \rangle =
2A\Delta t \, \delta_{i,k}\delta_{j,l}\delta_{m,n}\delta_{a,b} $ and then
letting \cite{Petschek83} $ \zeta_{ij}(n) = \frac{1}{\Delta
  x}[\nu^{(1)}_{i+1,j}-\nu^{(1)}_{i,j}+\nu^{(2)}_{i,j+1}-\nu^{(2)}_{i,j}]
$. In the simpler case of the SH--equation, $\zeta_{ij}(n)$ is a Gaussian
distributed field with zero mean and correlations $\langle
\zeta_{ij}(m)\zeta_{kl}(n) \rangle = 2A\Delta t \,
\delta_{i,k}\delta_{j,l}\delta_{m,n}$.

An inherent complication in this type of numerical simulations is the
conflicting constraints which the choice of the stepsizes is subject to.
The need for numerical accuracy requires $(\Delta x,\Delta t)$ to be
vanishingly small, whereas the finite computational power available
requires the opposite. Specifically, a linear stability analysis
\cite{Rogers88} of the above algorithm with the Laplacian given by
Eq.(\ref{La}) shows that, in order to avoid spurious solutions arising from
the subharmonic bifurcation, the dimensionless mesh size $\Delta x$ and
timestep $\Delta t$ must satisfy the relation $\Delta t <
2/[(\alpha-16/[3(\Delta x)^2])^2-\epsilon]$, where, as before,
$\alpha=1,1/2$ for the SH and DC systems, respectively.  In practice, the
size of $\Delta x$ is dictated by the the smallest length scale in the
problem, which is the selected wavelength $\lambda=2w$.  In order to avoid
lattice pinning it is desirable to have many lattice points per wavelength.
This quantity is given by $\lambda/\Delta x$, so by lowering $\Delta x$ any
number can be obtained.  However, from the above stability relation we see
that decreasing $\Delta x$ below unity drastically reduces the maximum
allowable size of the timestep, and hence increases the required computer
time.

We have performed our simulations using the values $(\epsilon,\Delta
x,\Delta t)=(0.25,2\pi/8,0.025)$ for the Swift--Hohenberg system and
$(\epsilon,\Delta x,\Delta t)=(0.05,1.0,0.05)$ for the diblock copolymer
system, where both set of values satisfy the appropriate stability
relations. In the SH system the selected wavelength is approximately
$2\pi$, so $\Delta x = 2\pi/8$ gives 8 lattice points per wavelength. The
corresponding quantity in the DC system, which we consider in the weak
segregation limit or small $\epsilon$, is approximately 9, since here the
selected wavelength is $\lambda=2\pi/(1/4-\epsilon)^{1/4}$.

Appropriate to a critical quench the systems were initially prepared in the
homogeneous single phase state by assigning to each lattice site a small
random number uniformly distributed about $\phi=0$.  Nonzero temperatures
were simulated using the fluctuation strengths $A=0.4$ and $A=0.1$ for the
SH and DC systems, respectively.

%
%----------------------------- Dynamical scaling ----------------------
%

%%%%%%%%%%%%%%%%%% FIGURE %%%%%%%%%%%%%%%%%%%%%%%
\begin{figure}[b]
  \begin{picture}(1,0.75)
    \put(-0.16,-0.95){\resizebox{1.25\columnwidth}{!}{\includegraphics{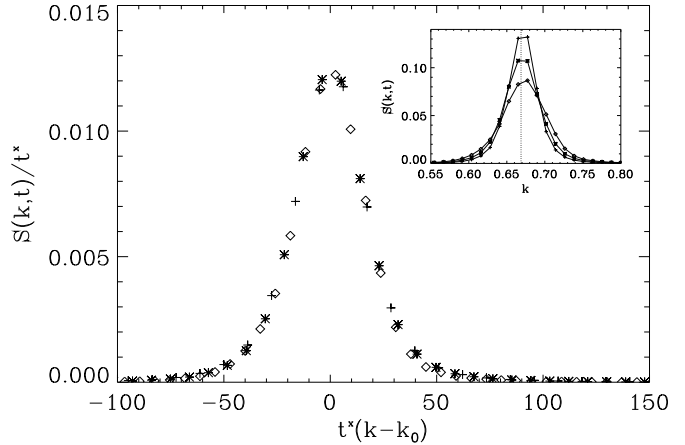}}}
  \end{picture}
  \caption{Test of the scaling form Eq.(\ref{Ssc}) and (inserted) 
    time evolution of the structure factor illustrated with data from
    simulations of the diblock copolymer system at zero thermal noise (Here
    depicted in arbitrary units). The scaling collapse was obtained with
    the value $x=1/5$ of the scaling exponent.  The data sets
    \{$\diamond$,$\ast$,$+$\} represent the (dimensionless) times
    \{$1.8\times 10^4$,$5.6\times 10^4$,$1.8\times 10^5$\}.}
  \label{fig:sscale}
\end{figure}
%%%%%%%%%%%%%%%%%%%%%%%%%%%%%%%%%%%%%%%%%%%%%%%%%
\subsection{Dynamical scaling}
We monitor the coarsening phenomenon by means of the usual structure factor
$S({\bf k},t)=|\phi_{\bf k}(t)|^2$, where $\phi_{\bf k}(t)$ is the Fourier
transform of the order parameter. The circularly averaged structure factor,
$S(k,t)$, is sharply peaked around the wavevector $k_0$ which corresponds
to the width of the rolls, and as time evolves it becomes increasingly
sharper and higher. Assuming dynamical scaling, the simplest scaling form
for the structure factor is
\begin{equation}
  \label{Ssc}
  S(k,t) = t^xf(t^x[k-k_0]),
\end{equation}
where $f(y)$ is a scaling function. This form implies that the width,
$\Delta k$, of the structure factor and its intensity, $S(k_0,t)$, scale as
$\Delta k \sim t^{-x}$ and $S(k_0,t)\sim t^x$.  In agreement with previous
work by a number of authors, our data from the SH system satisfies this
scaling form with the scaling exponents $x=1/5$ and $x=1/4$ at zero and
non-zero thermal noise respectively \cite{Elder92,CrossMeiron95,Hou97}.
Furthermore, we find that the diblock copolymer system also obeys
Eq.(\ref{Ssc}) with the same values of the scaling exponents
[Fig.(\ref{fig:sscale})].

%%%%%%%%%%%%%%%%%% FIGURE %%%%%%%%%%%%%%%%%%%%%
\begin{figure}
 \begin{center}
    \leavevmode
 \resizebox{\columnwidth}{!}{\includegraphics{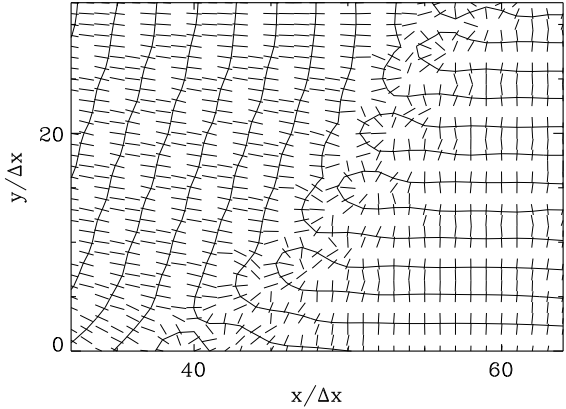}}
  \end{center}
%  \begin{picture}(1.0,0.5)
%    \put(-0.25,-0){\resizebox{1.2\columnwidth}{!}{\includegraphics{FIGURES/fig3.ps}}}
%  \end{picture}
  \caption{The local director field, ${\bf
      n}({\bf r})=\nabla\phi({\bf r})/|\nabla\phi(\bf r)|$, here
    illustrated as small bars, from which the correlation function
    Eq.(\ref{Cnn}) is computed. For visual clarity only directors near the
    domain boundaries (solid contours) are depicted.}
  \label{fig:df}
\end{figure}
%%%%%%%%%%%%%%%%%%%%%%%%%%%%%%%%%%%%%%%%%%%%%%%
A more direct method of probing the rolls increasingly orientational order
is computing a correlation function, $C_{nn}({\bf r},t)$, of the 'nematic'
order parameter ${\bf n} = \nabla\phi/|\nabla\phi|$, i.e. the unit vector
normal to surfaces of constant $\phi$ [Fig.(\ref{fig:df})].  Explicitly we
have computed the correlation function
\begin{equation}
  \label{Cnn}
  C_{nn}({\bf r},t) = \frac{2}{N^2} \sum_{\bf x}\langle 
[{\bf n}({\bf x}+{\bf
    r},t) \cdot {\bf n}({\bf x},t)]^2 \rangle - 1,
\end{equation}
where $N^2$ is the volume of the system and $\langle \ldots \rangle$ means
a statistical average implemented through several independent runs.  We
compute $\langle [{\bf n}({\bf r}_1)\cdot{\bf n}({\bf r}_2)]^2 \rangle$
rather than $\langle {\bf n}({\bf r}_1)\cdot{\bf n}({\bf r}_2) \rangle$
since we are interested only in the relative angle, $\theta({\bf r}_1,{\bf
  r}_2)$, between the directors at sites ${\bf r}_1$ and ${\bf r}_2$.  For
sites separated by large distances the corresponding directors can be
expected to completely decorrelated, thus $2\langle
\cos^2\theta\rangle_\theta -1 = 0$.

The time complexity of the algorithm for both $C_{nn}$ and its
corresponding structure factor, $S_{nn}({\bf k},t) = N^{-2}\sum_{\bf r}
C_{nn}({\bf r},t) \exp(i{\bf k}\cdot{\bf r})$, is $N^4$, and with $N=512$
excessive computer time is demanded. This problem can be circumvented by
introducing the two--dimensional tensor $Q_{ab}({\bf r},t) = n_a({\bf r},t)
n_b({\bf r},t)$, where $n_a$, $a=\{x,y\}$ are the components of ${\bf n}$.
In terms of $Q_{ab}$ Eq.(\ref{Cnn}) appears as
\begin{equation}
  \label{CnnQ}
   C_{nn}({\bf r},t) =  \frac{2}{N^2}  \sum_{\bf x} 
\langle Q_{ab}({\bf x}+{\bf
    r},t)Q_{ab}({\bf x},t)\rangle - 1,
\end{equation}
where summation over repeated indices is understood.  Since $S_{nn}$ now
has the form $S_{nn}({\bf k},t) = 2\langle Q_{ab}({\bf k},t)Q_{ab}(-{\bf
  k},t)\rangle - \delta_{{\bf k},{\bf 0}}$, $C_{nn}$ can quickly be
computed via $S_{nn}$ using Fast Fourier Transform \cite{NumRec92}.
%%%%%%%%%%%%%%%%%%% FIGURE %%%%%%%%%%%%%%%%%%%%%%%
\begin{figure}
 \begin{picture}(1.0,0.6)
     \put(-0.1,-.8){\resizebox{1.1\columnwidth}{!}{\includegraphics{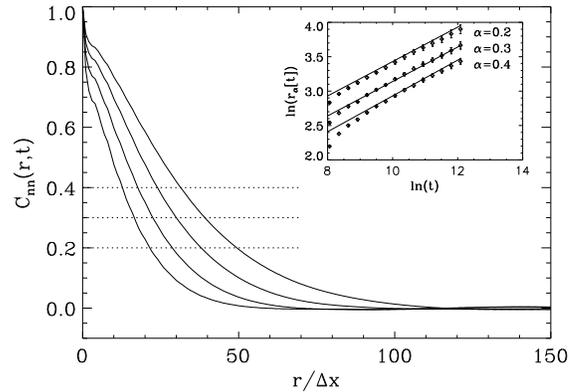}}}
    \end{picture}
  \caption{Time evolution of the director correlation 
    function Eq.(\ref{Cnn}) illustrated with four successive
    (dimensionless) times \{$5.6\times 10^3$,$1.8\times 10^4$,$5.6\times
    10^4$,$1.8\times 10^5$\} increasing from left to right. We extract the
    time evolution of the length scale, $L(t)$, by monitoring the
    $r_\alpha(t)$ for which $C_{nn}(r_\alpha(t))=\alpha$ where $\alpha<1$
    is some constant (the horizontal dotted lines show
    $\alpha=\{0.2,0.3,0.4\}$). The scaling exponent $y$ is extracted from a
    log-log plot of $r_\alpha(t)$ versus $t$ (inserted). The data shown
    result from a simulation of the SH system at zero thermal noise, and
    $y$ assumes the value $y=0.25\pm0.02$.}
  \label{fig:cscale}
\end{figure}
%%%%%%%%%%%%%%%%%%%%%%%%%%%%%%%%%%%%%%%%%%%%%%%%%%
The expected scaling form for the correlation function $C_{nn}$ is
\begin{equation}
  \label{FCnn}
  C_{nn}({\bf r},t) = F(r/L(t)),
\end{equation}
where $F$ is a scaling function, and $L$ is a length scaling as $L(t) \sim
t^y$. For both the Swift--Hohenberg and diblock copolymer system we find
this scaling form to be satisfied with the scaling exponents $y=0.25$ and
$y=0.30$ at zero and finite noise, respectively [Fig.(\ref{fig:cscale})].

The values of the scaling exponent $y$ agree with the findings of Hou et al
\cite{Hou97}. These authors measure the density, $\rho(t)$, of topological
defects in the Swift--Hohenberg system and find the algebraic decay
$\rho(t) \sim t^{-y}$ where $y=0.25$ and $y=0.30$ at zero and finite noise,
respectively. The boundaries between plane-wave domains consist of
topological defects. Therefore the defect density must scale as the
perimeter density of the domains which again scales as the reciprocal,
$L^{-1}$, of the linear size of the domains.  Furthermore, Hou et al find
that the energy of the Swift--Hohenberg system, Eq.(\ref{FSH}), decays as
the defect density. Also here the diblock copolymer system behaves as the
Swift--Hohenberg system.  Measuring the energy, as given by Eq.(\ref{FDC}),
we find the algebraic decay $E(t) \sim t^{-y}$ with the same values for $y$
as above.

%
%------------------------------ T H E O R Y -------------------------------
%

\section{Theory} \label{sec:Theory}
Theoretical analysis of the pattern dynamics of lamellar phases is
complicated by the presence of topological defects and current theories
apply only to systems without defects.  However, locally Type $I_s$ systems
exhibit nearly ideal lamellar structures where, in two dimensions, the
order parameter can be described as an amplitude
modulated plane wave
%\begin{equation}
%\label{modamp}
$\phi({\bf r},t) =[\phi_0A(x,y,t)e^{ik_0x}+c.c.]$,
%\end{equation}
where we have assumed lamellae perpendicular to the $x$--direction; $A$ is
a complex amplitude and $c.c.$ denotes the complex conjugate. Inserting
this form into the equations of motion, Eq.(\ref{SH}) or Eq.(\ref{DC}), we
obtain in the absence of noise the {\em Amplitude Equation}
\begin{equation}
\label{amp}
\tau_0\partial_tA=\epsilon A +\xi_0^2 
[\partial_x-(i/2k_0)\partial_y^2]^2A-g_0|A|^2A,
\end{equation}
where $\tau_0,\xi_0$ and $g_0$ are constants. The derivation of
Eq.(\ref{amp}) from the Swift--Hohenberg equation [Eq.(\ref{SH})] is
described in Ref.\cite{CrossHohenberg94} and the method of this reference
can easily be extended to the DC--equation [Eq.(\ref{DC})].  The amplitude
equation describes the dynamics of both the magnitude, $|A|$, and the
phase, $\theta({\bf r},t)$, of the complex amplitude $A$. By perturbing the
steady state solution of Eq.(\ref{amp}) we obtain, to lowest order in
$\epsilon$, the {\em Phase Equation} \cite{CrossHohenberg94}
\begin{equation}
\label{phase}
\partial_t \theta = D_\parallel 
\partial_x^2\theta+D_\perp\partial_y^2\theta,
\end{equation}
where $D_\parallel$ and $D_\perp$ are diffusion coefficients in the
parallel and normal directions, respectively.  Dimensional analysis of
Eq.(\ref{phase}) implies a $t^{1/2}$--growth of the characteristic length
scale, in disagreement with numerical investigations which favour a smaller
value of the growth exponent. However, as discussed below, by considering
how a curved interface relaxes, working to second order in $\epsilon$, a
transient regime with $t^{1/4}$--growth can be predicted.

%
%---------------------Projection operator method-----------------
%
\subsection{Projection operator method}
In order to follow the slowly varying orientation of the rolls (or
lamellae) Elder et al \cite{Elder92} introduces a coordinate
system that tracks the interface given by the points at which $\phi=0$.
Specifically the Cartesian coordinates $x$ and $y$ are mapped onto
curvilinear coordinates $(u,s)$, where $u$ and $s$ are locally normal and
parallel to the lines $\phi({\bf r},t)=0$.  Assuming that the curvature of
the individual rolls are small, the Laplacian in the new coordinates
becomes
\begin{equation}
  \label{Lacurv4}
  \nabla^2 \simeq \frac{\partial^2}{\partial u^2}+
\kappa\frac{\partial}{\partial u} + 
\frac{\partial^2}{\partial s^2},
\end{equation}
where $\kappa$ is the local curvature.  Assuming that the stationary
solution of the one--dimensional Swift--Hohenberg equation is a good
approximation in the normal direction, Eq.(\ref{SH}) becomes
\begin{equation}
  \label{SHstat}
  \frac{\partial \phi^s}{\partial u}\frac{\partial 
u}{\partial t} =
  2\kappa\left( \frac{\partial\phi^s}{\partial u}+
  \frac{\partial^3\phi^s}{\partial u^3}\right) + 
\kappa_{ss}
  \frac{\partial\phi^s}{\partial u} + \Delta, 
\end{equation}
where $\kappa_{ss}=\partial^2\kappa/\partial s^2$ and $\phi^s$ is the
solution of $ \phi^s(u({\bf r},t))^3
=[\epsilon-(k_0^2+\partial_u^2)^2]\phi^s(u({\bf r},t))$. The final term,
$\Delta$, in Eq.(\ref{SHstat}) contains terms of higher order in $\kappa$,
and terms involving the derivative of $\kappa$ in the direction normal to
the lamellae: $\Delta = (\kappa_{uu} + \kappa\kappa_u)\partial_u\phi^s +
(2\kappa_u + \kappa^2)\partial_u^2\phi^s$, where $\kappa_u$ means
$\partial_u\kappa$, etc. Application of the projection operator,
\begin{equation}
\label{proj}
\frac{k_0}{2\pi}\int_{-\pi/k_0}^{\pi/k_0}du\,\partial_u\phi^s, 
\end{equation}
to Eq.(\ref{SHstat}) produces the final result
\begin{equation}
  \label{SHv}
  v = -a\kappa+\kappa_{ss},
\end{equation}
where $v=\partial_tu$ is the interface velocity,
$a=-2(k_0^2+\beta/\sigma)$, $ \sigma=(k_0/2\pi)\int_{-\pi/k_0}^{\pi/k_0}
du\,(\partial_{u}\phi^s)^2$, and $\beta =
(k_0/2\pi)\int_{-\pi/k_0}^{\pi/k_0} du\,
(\partial_{u}\phi^s)(\partial_{u}^3\phi^s)$.  The term involving $\Delta$
drops out from the final result, because $\Delta\,\partial_u\phi^s$ can be
written as $\partial_u [(\kappa_u + \kappa^2/2)(\partial_u\phi^s)^2]$,
which vanishes when integrated over one lamellar thickness.

In order to evaluate the coefficient $a$ in Eq.(\ref{SHv}) the stationary
solution is expanded to leading order in $\epsilon$, yielding
\cite{PomeauMan79}.
\begin{equation}
\label{statex}
\phi^s(u)=\Phi_1\cos(k_0u)+\Phi_3\cos(3k_0u),
\end{equation}
with coefficients $\Phi_1=\sqrt{4\epsilon/3}$ and $\Phi_3=-\Phi_1^3/256$.
Using this expansion we find $a=\epsilon^2/256$, remembering that
$k_0^2=1$.

Applying the same analysis as above to the DC--equation, the equation
corresponding to Eq.(\ref{SHstat}) becomes
\begin{equation}
  \label{DCstat}
  \frac{\partial\phi^s}{\partial u}\frac{\partial u}{\partial t} =
  \kappa\left(\frac{\partial \phi^s}{\partial u}+2\frac{\partial^3
    \phi^s}{\partial u^3}-3(\phi^s)^2\frac{\partial\phi^s}{\partial
    u}\right)\! +\! \kappa_{ss}\frac{\partial\phi^s}{\partial u} \!+ \!\Delta
\end{equation}
where $\phi^s$ is the solution of $\partial_u^2\phi^s(u({\bf r},t))^3
=[\epsilon-(1/2+\partial_u^2)^2]\phi^s(u({\bf r},t))$, and $\Delta$ has the
same meaning as in Eq.(\ref{SHstat}). Using the projection operator
[Eq.(\ref{proj})] we retrieve Eq.(\ref{SHv}), only now with
$a=-(1+2\beta/\sigma-3\gamma/\sigma)$ where
$\gamma=(k_0/2\pi)\int_{-\pi/k_0}^{\pi/k_0}
du\,(\phi^s)^2(\partial_u\phi^s)^2,$ and $\sigma$ and $\beta$ are as
defined above. The quantities $\sigma$, $\beta$, and $\gamma$ can be 
determined by substituting the form Eq.(\ref{statex}) into the free-energy
functional Eq.(\ref{FDC}), and minimizing with respect to $k_0$, $\Phi_1$, and 
$\Phi_3$. To the required order in $\epsilon = 1/4 - \Gamma$, the result is 
$k_0 = \Gamma^{1/4} = (1/4 - \epsilon)^{1/4}$, $\Phi_1^2 = (8/3)\epsilon + 
(19/6)\epsilon^2$, and $\Phi_3 = -(9/128)\Phi_1^3$, leading (after some
algebra) to $a = (45/32)\epsilon^2$, correct to leading non-trivial order 
in $\epsilon$. 

In this approximation $a$ is a very small number, $a \simeq 2.4 \times
10^{-4}$ and $a \simeq 3.5 \times 10^{-3}$ in the SH and DC systems when
$\epsilon=0.25$ and $\epsilon=0.05$, respectively.  Dimensional analysis of
Eq.(\ref{SHv}) therefore implies a crossover in the growth of the
characteristic length scale from a transient $t^{1/4}$--growth to an
asymptotic $t^{1/2}$--growth.  The crossover occurs approximately when
$(at)^{1/2}=t^{1/4}$, that is, when $t \simeq 1.7 \times 10^7$ in the SH
system and when $t \simeq 8 \times 10^4$ in the DC system.  These
crossover times far exceed the latest times we have been able to probe in
our simulations, but though (as pointed out by Elder et al \cite{Elder92})
an appealing interpretation of the numerical results is that they witness
the transient regime, there is no numerical evidence of any crossover
behaviour. However, since the estimated crossover time for the DC system is
three orders of magnitude smaller than the corresponding time in the SH
system, the DC system is the obvious candidate for future investigations.

%
%---------------------Relaxation of a modulated interface --------------
%

\subsection{Relaxation of a modulated interface}
The same problem can be investigated using a general approach to growth
exponents recently developed by one of the authors \cite{Bray98}.

\begin{figure}
 \begin{picture}(1.0,0.6)
      \put(-0.33,-1.2){\resizebox{1.5\columnwidth}{!}{\includegraphics{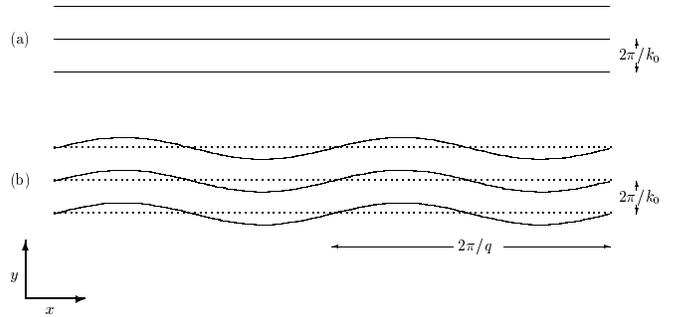}}}
    \end{picture}
\caption{We here sketch (a) the unperturbed lamellar phase and (b) a
    modulated lamellar phase. The perturbation
    $\tilde{\phi}=A(t)\cos(qy)\psi_q$ introduce spatial variations in the
    pattern at the length scale $L=2\pi/q$.}
  \label{fig:relax}
\end{figure}

We consider a small regular perturbation of the perfect lamellar phase
 and wish to determine the rate at which the system
relaxes to its ground state.  Setting $\phi(x,y,t)=\phi_L(x)
+\tilde{\phi}(x,y,t)$, where $\phi_L(x)$ is the stationary lamellar
solution of the Swift--Hohenberg equation [Eq.(\ref{SH})],
\begin{equation}
\label{lamel}
0 = \epsilon \phi_L -(k_0^2+\partial_x^2)^2 \phi_L -\phi_L^3,
\end{equation}
and $\tilde{\phi}$ is a small perturbation, the linearized equation of
motion for $\tilde{\phi}$ becomes
\begin{equation}
  \label{utilde}  
\partial_t\tilde{\phi}=\epsilon\tilde{\phi}-(k_0^2+\nabla^2)^2\tilde{\phi}-
3\phi_L^2 \tilde{\phi}.
\end{equation}
A modulation of the lamellar phase with wave vector $q \ll k_0$ is
$\phi(x,y,t) = \phi_L[x+A(t)\cos(qy)] \simeq
\phi_L(x)+\phi_L'(x)A(t)\cos(qy)$, where $\phi_L'$ means $\partial_x\phi_L$
and the amplitude, $A$, of the modulation is assumed small compared to the
lamellar spacing [Fig.(\ref{fig:relax})]. More generally, we can write the
modulated phase as
\begin{equation}
  \label{mod}
  \phi(x,y,t) = \phi_L(x)+A_0e^{-\omega_q t}\cos(qy) 
\psi_q(x)
\end{equation}
where the function $\psi_q(x)$ has the same periodicity as $\phi_L$,
($\phi_L \sim \cos(k_0x), k_0=1$), and satisfies the condition
$\psi_{q=0}(x)=\phi_L'(x)$, since $q=0$ represents a uniform translation of
the lamellae. We have written the time dependence of the amplitude as
$A(t)=A_0 e^{-\omega_q t}$, describing the decay of the eigen--perturbation
of wave vector $q$ at a characteristic rate $\omega_q$.

Insertion of Eq.(\ref{mod}) into Eq.(\ref{utilde}) yields an eigenvalue
equation, $\hat{H}\psi_q=\omega_q\psi_q$, for the relaxation rate with
'Hamiltonian'
\begin{equation}
  \hat{H} =(k_0^2+\partial_x^2)^2+3\phi_L^2(x)-\epsilon -
  2q^2(\partial_x^2+k_0^2)+q^4.
\end{equation}
The $q=0$ mode corresponds to an uniform displacement of the interface
wherefore $\omega_{q=0}$ is fixed to zero. Since $\psi_0=\phi_L'$, this
condition corresponds to
\begin{equation}
\label{cond}
0= \epsilon \phi_L'-3\phi_L^2\phi_L'-\phi_L'-2\phi_L'''-\phi_L''''',
\end{equation} 
which is satisfied because Eq.(\ref{cond}) is the derivative of
Eq.(\ref{lamel}).
%%%%%%%%%%%%%%%%%%%%%% FIGURE %%%%%%%%%%%%%%%%%%%%%%%%%%%%%
%%%%%%%%%%%%%%%%%%%%%%%%%%%%%%%%%%%%%%%%%%%%%%%%%%%%%%%%%%%
In the limit $\epsilon \rightarrow 0$ we have $\phi_L \rightarrow 0$, so
for $\epsilon=0$ the eigenvalue equation reads $ -\omega_q \psi_q =
-(k_0^2-q^2)^2\psi_q -2(k_0^2-q^2)\psi_q''-\psi_q''''.  $ Remembering that
$\psi_q$ is periodic in $x$ with wavevector $k_0$, this implies
$\omega_q=q^4$. For small $\epsilon$ we expect $\omega_q=a q^2+q^4$, where
$a$ is a constant. We can verify this assertion and determine the value of
$a$ by treating the $q$--dependent part of $\hat{H}$ as a perturbation, and
calculate the first-order correction to the ground state eigenvalue using
standard perturbation techniques. The unperturbed eigenfunction for
$\hat{H}$ is $\psi_0=\phi_L'$ with eigenvalue zero.  Obviously the
$q^4$-term in the perturbation just gives a contribution $q^4$ to the
eigenvalue. The $O(q^2)$ contribution is
$$ aq^2=-2q^2\frac{\int 
dx\,\phi_L'(\partial_x^2+k_0^2)\phi_L'}{\int
  dx\,(\phi_L')^2}=2q^2\left[\frac{\int 
dx\,(\phi_L'')^2}{\int
  dx\,(\phi_L')^2}-k_0^2\right],
$$ and with $\phi_L$ expanded as previously [Eq.(\ref{statex})] we have
$a=\epsilon^2/256$. Thus the relaxation rate is
\begin{equation}
  \label{omegak}
  \omega_q = \frac{\epsilon^2}{256}q^2+q^4.
\end{equation}
We notice that the coefficient $a$ here assumes the same value as
determined above by the projection operator method, and that a dimensional
analysis of Eq.(\ref{omegak}) thus predicts the same crossover behaviour as
did the analysis of Eq.(\ref{SHv}) for the interface velocity. Furthermore,
we notice that the formal expression for the coefficient $a$ is identical
to that obtained from the projection operator method, as may be seen from
an integration by parts, i.e., the result holds generally, not just to the
order given by the expansion Eq.(\ref{statex}).

Due to the more complicated structure of the DC--equation [Eq.(\ref{DC})] a
similar analysis of interfacial relaxation in diblock copolymers has not
yet proved possible. The main difficulty is that the `Hamiltonian' operator 
$\hat{H}$ for this case is not self-adjoint, even for $q=0$, with the 
result that a perturbative calculation of $\omega_q$ requires not only  
the null eigenfunction $\phi_L'(x)$ of the $q=0$ operator $\hat{H}_0$, 
but also the null eigenfunction of the adjoint operator $\hat{H}^\dagger_0$, 
which we have so far been unable to determine.  

%
%-------- S U M M A R Y--A N D--D I S C U S S I O N------------------
%

\section{Summary and Discussion} \label{sec:Discussion}

By numerical investigations we have found evidence of identical coarsening
dynamics for the lamellar phase of the Swift--Hohenberg and diblock
copolymer systems. This suggest that both systems belong to the same
universality class. We have extracted temperature dependent dynamical
scaling exponents for the characteristic length scale partly by computing
the ordinary structure factor and partly by computing a correlation
function [Eq.(\ref{Cnn})] of the director field.  Surprisingly the two
methods yield different scaling exponents indicating that the scaling
phenomenon in question is non--trivial. We have no good understanding of
the reasons for this discrepancy, but it should be noticed that the length
scale extracted from the structure factor does not have the same immediate
geometrical interpretation as has the length scale extracted from the
director-field correlation function. Furthermore, the fact that the length
scale, $L_{nn}$, arising from the director field correlation function
scales with the same growth exponents as the energy suggests that $L_{nn}$
is the physically important length scale in the system.

Theoretically, by considering how curved interfaces relax we have
demonstrated that the projection operator method, when applied to either of
the two systems, results in the same scaling exponents.  This finding
supports the suggestion from our numerical work that the coarsening
dynamics of the Swift--Hohenberg and diblock copolymer systems belong to
the same universality class. However, the theoretical analysis applies only
to defect free systems and does not explain the observed
temperature-dependence of the growth exponents. A thorough understanding of
the coarsening phenomenon here considered requires a theoretical treatment
which successfully includes the simultaneous effects of both interfacial
relaxation and defect--defect interactions.

\subsection*{Acknowledgements}

J.J.C. wish to thank the Theory Group of the Schuster Laboratory, The
University of Manchester, for hospitality during a nine month visit.

%%REFERENCES

%%FIGURES:

%

%
%

\end{document}